\documentstyle[aps,floats,epsfig,prb]{revtex}
\begin{document}
\draft
\title{Strong-coupling perturbation theory for the two-dimensional 
Bose-Hubbard model in a magnetic field}
\author{M. Niemeyer$^{\dagger}$, 
J. K. Freericks$^*$, and 
H. Monien$^{\dagger}$}
\address {$^{\dagger}$Physikalisches Institut der Univerit\"at Bonn, 
Nu\ss allee 12, D-53115 Bonn, Germany\\
$^*$Department of Physics, Georgetown University, Washington, D.C. 20057}

\date{\today}
\maketitle
\widetext
\begin{abstract}
  The Bose-Hubbard model in an external magnetic field is investigated
  with strong-coupling perturbation theory.  The lowest-order secular
  equation leads to the problem of a charged particle moving on a
  lattice in the presence of a magnetic field, which was first treated
  by Hofstadter.  We present phase diagrams for the two-dimensional
  square and triangular lattices, showing a change in shape of the
  phase lobes away from the well-known power-law behavior in zero
  magnetic field.  Some qualitative agreement with experimental work
  on Josephson-junction arrays is found for the insulating phase
  behavior at small fields.
\end{abstract}
\pacs{PACS numbers: 05.30.Jp, 05.70.Fh, 67.40.Db}
\tighten
\section{Introduction}

The simplest model of strongly interacting bosons is the Bose-Hubbard
model (BHM), which has been used to describe superfluid
helium\cite{scal94}, Cooper pairs in thin granular superconducting
films\cite{jaeger89liu91} and Josephson junction
arrays\cite{rod9395}. Much theoretical work has concentrated on
the phase diagram of the BHM in zero magnetic field\cite{fisher89}
because of the technical problems associated with introducing an
external magnetic field (such as the sign problem in quantum Monte
Carlo simulations).  Experimentalists on the other hand have
concentrated on studying systems in an external magnetic field because
the phase transition can be tuned by adjusting the magnetic field
rather than changing the samples measured\cite{mooij,pannetier84}.  We
study the BHM on two-dimensional lattices in a perpendicular magnetic
field by extending the strong-coupling perturbation theory for the
field-free case\cite{monien}. This theoretical technique can
incorporate magnetic-field dependence in a straightforward manner and
is useful in studying field-tuned
transitions.  We concentrate on pure systems in this contribution and do not
include any effects due to disorder.
We present zero-temperature phase diagrams, study
excitation-gap energies, calculate the dynamical critical exponent
$z\nu$ for small fields, and compare our theoretical results with
experimental ones.

The BHM contains the key physics of a many-particle bosonic system
with competing potential and kinetic energy effects.  The typical
zero-temperature phase diagram for the non-magnetic case shows
incompressible Mott-insulating (MI) phases surrounded by 
compressible superfluid phases (SF) \cite{fisher89}. The insulator to
superfluid transitions at the tip of the lobes (where the density
remains constant) are driven by quantum phase fluctuations, while
those at the sides of the lobes (where the density varies) are driven by density
fluctuations, i.e. particle or hole excitations. Introducing a
magnetic field is expected to increase the region of the MI phase
because the localizing effect on the itinerant bosons reduces the
stability of the SF phase.

We consider bosons with total spin $0$. The only effect of a perpendicular
magnetic field $\vec{H}$ is then on the orbital motion of the bosons, 
which effects changes in the phase of the hopping-matrix
$\hat{T}=(t_{jk})$ between lattice sites $j$ and $k$. By choosing a
Landau-gauge for the vector-potential $\vec{A}(\vec{r}) = H
\hspace{0.02cm} (0,x,0)$, the Bose-Hubbard Hamiltonian in an external
magnetic field becomes
\begin{equation} 
  H_{BHM} =- \sum_{<jk>} ( t_{jk}^{\phantom{\dagger}}
  b^{\dagger}_{j} b^{\phantom{\dagger}}_{k}
  + {\it{h.c.}}
  ) + \frac{U}{2} \sum_{j} 
  \hat{n}_{j} (\hat{n}_{j}-1)
  - \mu \sum_{j} \hat{n}_{j},
\end{equation}
where the hopping matrix is nonzero only between 
nearest neighbors and is given by
\begin{equation}
  t_{jk} = t e^{-i 2 \pi \tilde{A}_{jk}} ,
  \hspace{0.4cm}
  \tilde{A}_{jk} =  \frac{1}{\phi_{0}}\int_{j}^{k} 
  \vec{A}(\vec{r})\cdot d\vec{r}.
\end{equation}
This hopping matrix is Hermitian because $t$ is real and $
\tilde{A}_{jk} = -\tilde{A}_{kj}$.  Here the boson creation operator
for lattice site $j$ is $b^{\dagger}_{j}$,
$\hat{n}_{j}=b^{\dagger}_{j} b^{\phantom{\dagger}}_{j}$ is the
corresponding number operator, $U$ is the on-site repulsion of the
bosons, and $\mu$ is the chemical potential. We choose $U$ to be our energy
scale and measure all energies in units of $U$.
The magnetic flux quantum is given by $\phi_{0}=\frac{h
  c}{e}$, and the magnetic flux per plaquette $2\pi\phi =
\oint\vec{A}(\vec{r})\cdot d\vec{r}\sim a^2 H$ is a measure of the strength
of the magnetic field $\vec{H}$ (where $a$ is the lattice spacing).

The form of the zero-temperature phase diagram can be understood by
starting from the atomic limit\cite{monien}, where the hopping $t$ 
is zero and
every site is occupied by a fixed number of bosons $n_0$.  The energy
to add one boson onto a site with $n_0$ bosons
is $E(n_0+1)-E(n_0)=n_0U-\mu$, so that
there is a finite energy gap when $(n_0-1)U<\mu < n_0U$ and the system
is an incompressible Mott insulator.  When $\mu=n_0U$, then all states
with a density between $n_0$ and $n_0+1$ bosons per site are
degenerate in energy, and the system becomes a compressible fluid. As
the strength of the hopping matrix elements increases, the range of
the chemical potential about which the system is incompressible
decreases.  The Mott insulator phase disappears at a critical value of
the hopping matrix elements (which depends on the strength of the
magnetic field and the density of the insulating phase) 
and the system becomes a superfluid (see
Fig.\ref{phasesqtri}).  We provide a systematic study of the BHM in
a magnetic field by examining the system in a perturbative expansion
about the atomic limit with the boson kinetic energy acting as the
perturbation.  

The paper is organized as follows: Section II describes the formalism
used in the strong-coupling perturbation theory in the presence of a
magnetic field.  Section III presents the results for the phase
diagrams and the excitation energies in a magnetic field, and Section
IV contains the conclusions.

\section{Formalism}

Our procedure is to calculate the ground-state energy of the MI phase
with $n_0$ bosons per site $E_g(n_0,t)$, and of the excited states in
the charge sector with one extra boson $E_{p}(n_0,t)$ and one extra
hole $E_{h}(n_0,t)$, in a Rayleigh-Schr\"odinger perturbative
expansion in the hopping matrix element $t$ (the kinetic energy
is chosen as the perturbative part of the Hamiltonian).  When the
energy of the MI and the state with one extra boson are equal, the
system undergoes a phase transition from the incompressible MI phase
to the compressible SF phase with density larger than $n_0$.  The
similar occurs when the state with one extra hole is degenerate with
the MI phase (except now the density of the SF phase is less than
$n_0$).  The detailed formalism of the strong-coupling expansion for the
ground-state energies has already been presented\cite{monien}. The only 
modifications of the
previous calculations needed here are to take into account the fact
that the hopping matrix now has a complex phase and the changes required for
the nonbipartite hopping matrix of the triangular lattice.  The important 
parameter that enters
the results is the minimal eigenvalue $\epsilon t$ of the kinetic energy matrix
$-t_{jk}$ which includes a factor $\epsilon$ that
depends on the magnetic field.  This parameter determines how the
degeneracy is lifted in the first-order secular equation for the
energy of the excited states in the charge sector.  
Formally this solution of the minimal
eigenvalue is identical to finding the band minimum in
the Hofstadter problem\cite{hofstadter76}.

The Mott phase diagram is determined by the two Mott phase boundaries---one 
for the particle excitations [where $E_{p}(n_0,t)-E_g(n_0,t)=0$]
and one for hole excitations [where $E_{h}(n_0,t)-E_g(n_0,t)=0$]. For
each value of $t$ there is a critical value of the chemical potential
where the system changes phase from an incompressible to compressible
fluid.  The upper and lower curves for the Mott phase lobe are then
described by this critical value of the chemical potential
$\mu_{p/h}(t)$.  The results of our expansion through third-order are
\begin{eqnarray}
  \label{mueq}
   \mu_{p}(t)&=&n_0+\epsilon(n_0+1)t-n_0(n_0+1)\epsilon^2 t^2+\frac{n_0}{2}
                    (5n_0+4)zt^2\nonumber\\
             &-&n_0(n_0+1)[(2n_0+ 1)(-2-\epsilon^2)+z(\frac{25}{4}n_0
                  +\frac{7}{2})]\epsilon t^3\nonumber\\
             &+&12\delta_{lat}n_0(\frac{31}{4}n_0^2+\frac{21}{2}n_0+3) 
                  \cos(2\pi\phi)t^3+O(t^4)\nonumber\\
   \mu_{h}(t)&=&n_0-1-\epsilon n_0 t+n_0(n_0+1)\epsilon^2 t^2-
                    \frac{n_0+1}{2}
                    (5n_0+1)zt^2\nonumber\\
             &+&n_0(n_0+1)[(2n_0+1)(-2-\epsilon^2)+z(\frac{25}{4}n_0
                  +\frac{11}{4})]\epsilon t^3\nonumber\\
             &-&12\delta_{lat} (n_0+1)(\frac{31}{4}n_0^2+5n_0+\frac{1}{4}) 
                  \cos(2\pi\phi)t^3+O(t^4), 
\end{eqnarray}
where $\delta_{lat}=0$ or $1$ for the square or the triangular
lattice, respectively, and $z$ is the corresponding 
number of nearest neighbors (4
or 6, respectively).  These results have been verified by both
analytical small-cluster calculations and by numerical
cluster expansions.  

The magnetic field appears in two places: (1) The magnetic field couples
to the orbital current of the bosons as the particle or hole travels
around a lattice plaquette and encloses the flux $2\pi\phi$. 
In our third-order
calculation, the orbital coupling only enters for the triangular
lattice (it enters at fourth order for the square lattice as
four hops are required to enclose a
plaquette). (2) The other effect of the magnetic field is to
change the minimal energy of the extra particle or hole moving in the Mott
phase background.
Although the location of this mininum in the Brillouin zone 
is gauge-dependent, its value is not (thus 
$\epsilon = min_{\vec{k}\in\relax{\rm I\kern-.18em B}}\epsilon(\vec(k))$).

We consider a rational flux $2\pi\phi=2\pi p/q$.  To find the minimal 
eigenvalues $\epsilon t$
of the hopping matrix $\hat{T}$, we follow Bellisard \cite{kreft90}
and Hasegawa\cite{rice89}. In the Landau gauge, the system maintains its
translational invariance in the $x$-direction while it requires $m$ steps for
translational invariance in the $y$-direction.
For the square lattice, the period $m$ equals
$q$, for the triangular lattice, $m=q/2$ for even $q$ and $m=q$ for
odd $q$. A Fourier-transformation now changes $\hat{T}$ to the
following $m\times m$-matrix $\tilde{T}_{\phi}(\vec{k})$:
\begin{equation}
\tilde{T}_{\phi}(\vec{k}) = t \left(
\begin{array}{ccccc}
M_1 & A_1 & 0 & & A_m^{\star} \\
A_1^{\star} & M_2 &\ddots &\ddots&  \\
   0&\ddots&\ddots&\ddots& 0\\
    &\ddots&\ddots&M_{m-1}&A_{m-1}\\
A_m & &0& A_{m-1}^{\star} & M_m 
\end{array} 
\right) ,
 \label{thematrix}
\end{equation}
with
\begin{eqnarray}
  M_n^{sq}  &=& - 2 \cos(k_y a+2\pi n\phi),\quad 
  A_n^{sq}  = - e^{i k_x a},\\
  M_n^{tri} &=& - 2 \cos(k_y a + 4\pi n\phi), \quad
  A_n^{tri} = - e^{i k_x a} (1+e^{i 2\pi\phi (2n+1)+i k_y a}).
\end{eqnarray}
The eigenvalues within the corresponding Brillouin-zone $\relax{\rm
I\kern-.18em B}= \{ \vec{k}\hspace{0.1cm}|\hspace{0.1cm} 0\leq k_x
\leq 2\pi/m, 0\leq k_y\leq 2\pi/m\} $ determine the $m$ energy bands
of a boson moving in the magnetic field.
\begin{figure}[t]
  \begin{center}
    \protect\epsfig{file=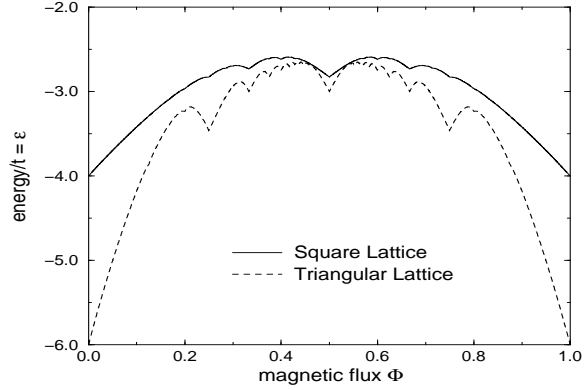,height=6cm,width=8.6cm,angle=0}
    \protect\caption[]{Band minimum of the 
       magnetic band structure $\epsilon(\phi)$ for the 
    square and triangular lattice.}
    \label{butterfly}
  \end{center}
\end{figure}
The minimal eigenvalue of the band-structure
$\epsilon t$ is shown in Fig.~\ref{butterfly} as a function of the magnetic 
flux per plaquette $2\pi\phi$. This is the parameter that enters
Eq.~(\ref{mueq}) to determine the Mott phase boundary.
Notice how on the square lattice the largest dip is at $\phi=1/2$ 
followed by 1/3, 1/4, and 2/5, ...
while on the triangular lattice the sequence corresponds to
1/2, 1/4, 1/3, and 3/8, ...
The relation $\epsilon(\phi)=\epsilon(1-\phi)$
holds for all lattices, since the flux of $2\pi$ is equivalent to a
flux of 0.  Hence, the maximal magnetic field that can be applied
corresponds to $\phi=1/2$.  This maximal field configuration 
is realized by all real hopping
matrix elements for each lattice: on the square lattice one takes
three positive and one negative matrix element on each plaquette, while on the
triangular lattice one takes all matrix elements to be negative.  In
particular, the ``fully frustrated'' case of $\phi=1/2$ on a
triangular lattice is the only nonzero magnetic-field case that can be
easily treated by a high-order expansion\cite{monien98} because it
maintains the full periodicity of the triangular lattice in zero
magnetic field.
 
Rather than plotting the phase diagrams for just a third-order
expansion, we choose to extrapolate our results using knowledge about
the overall structure of the phase diagram.  There are numerous ways
in which one can envision extrapolating the results of our third-order
expansion to higher order. It has been demonstrated by Elstner and
Monien that a Pade analysis of the strong coupling perturbation series
yields rapid convergence in zero field \cite{monien98}.  
We apply this method to the magnetic-field case for small magnetic fields
by using a Pade approximant\cite{guttmann} for the logarithmic
derivative of the difference in particle and hole Mott phase boundaries
$\Delta (t)=\mu_p(t)-\mu_h(t)$.  We assume the same behavior as found in
the zero-field case, where the tip of the Mott lobe has a power law
``critical point'' with an exponent $z\nu$, $\Delta(t)=A(t)(t_{crit}-t)^{z\nu}$.
Then the Pade analysis fits
\begin{equation}
\label{znuformel}
\frac{\partial}{\partial t}
\log{\Delta(t)}=\frac{z\nu}{t-t_{crit}} + \frac{A'(t)}{A(t)} ,
\end{equation}
to estimate the critical point and the dynamical critical exponent
(the pole determines $t_{crit}$ and the residue determines $z\nu$).
The rest of the Pade approximant determines $A(t)$, which then allows
$\Delta(t)$ to be constructed.  A similar Pade analysis for the
midline of the Mott lobe
$\mu_m(t)=\frac{1}{2} [\mu_p(t)+\mu_h(t)]$ (which is a regular
function of $t$) then allows the entire phase diagram to be
constructed.

\section{Results}

Atomic systems with a lattice spacing $a$ of around $2
A^{\hspace{-0.14cm}^{o}}$ require a field of $H\cong 5\times
10^3$ Tesla for a half flux quantum per plaquette. 
This is too large a field to be accessed experimentally,
hence atomic systems always lie in the low-field region, where
the perturbation theory is most accurate.  However for
macroscopic lattice systems, with $a\cong 2\times 10^{-4} cm$ (as in a
Josephson-junction array) the whole magnetic field range is attainable by
experiments with fields as low as $H\cong 0.5$ Gauss. We show below how
our theoretical results compare to the 
superfluid-insulator transitions on two dimensional 
Josephson-junction arrays \cite{mooij}.
\begin{figure}[t]
\setlength{\unitlength}{1mm}
\def\hoehe{64mm}
\def\breite{64mm}
\vspace{1.0cm}
\begin{picture}(60,50)(0,0)
\put(21,0){
        \makebox{ \epsfig{file=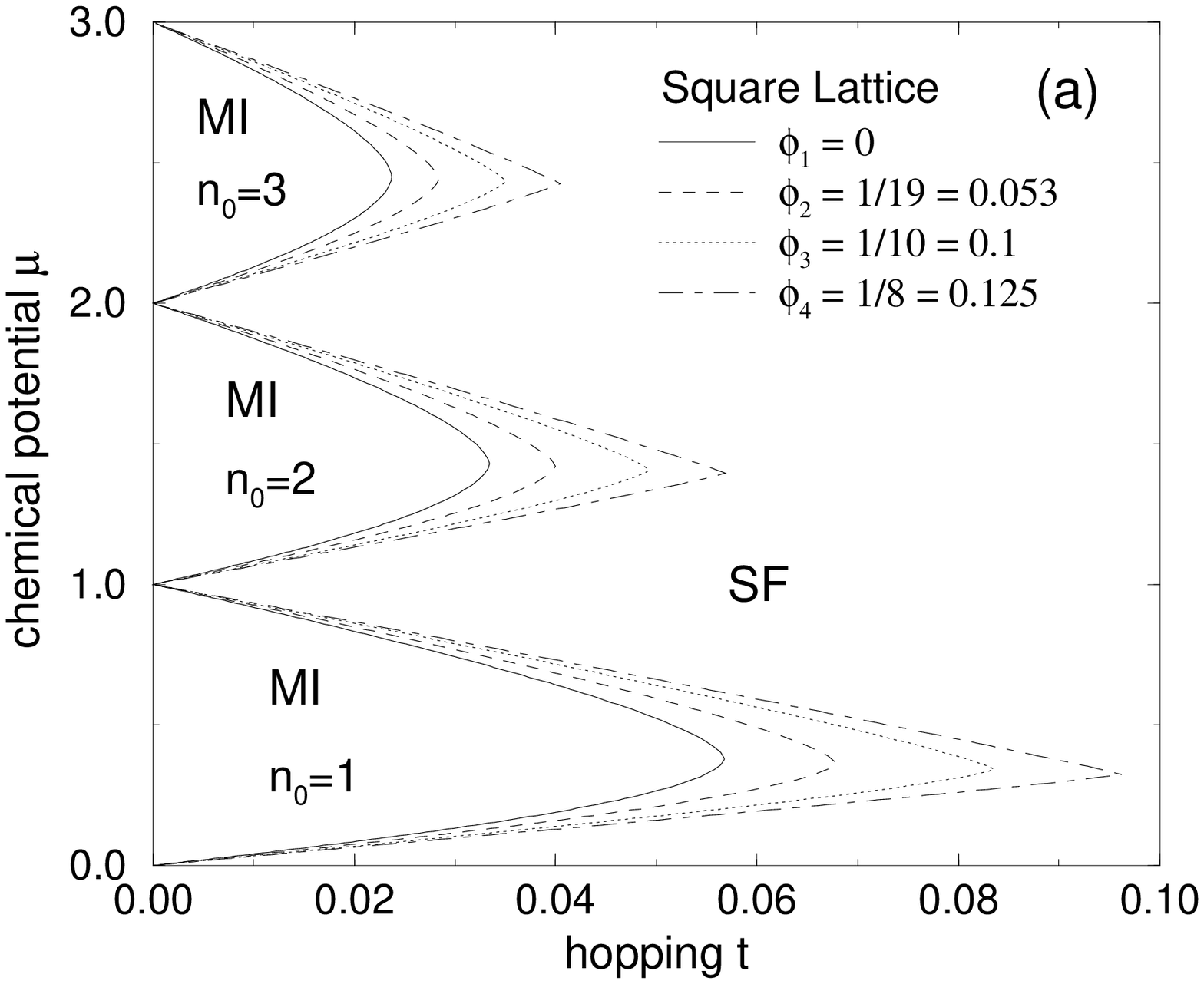,
                  height=\hoehe,width=\breite}}
        }
\put(86,0){
        \makebox{ \epsfig{file=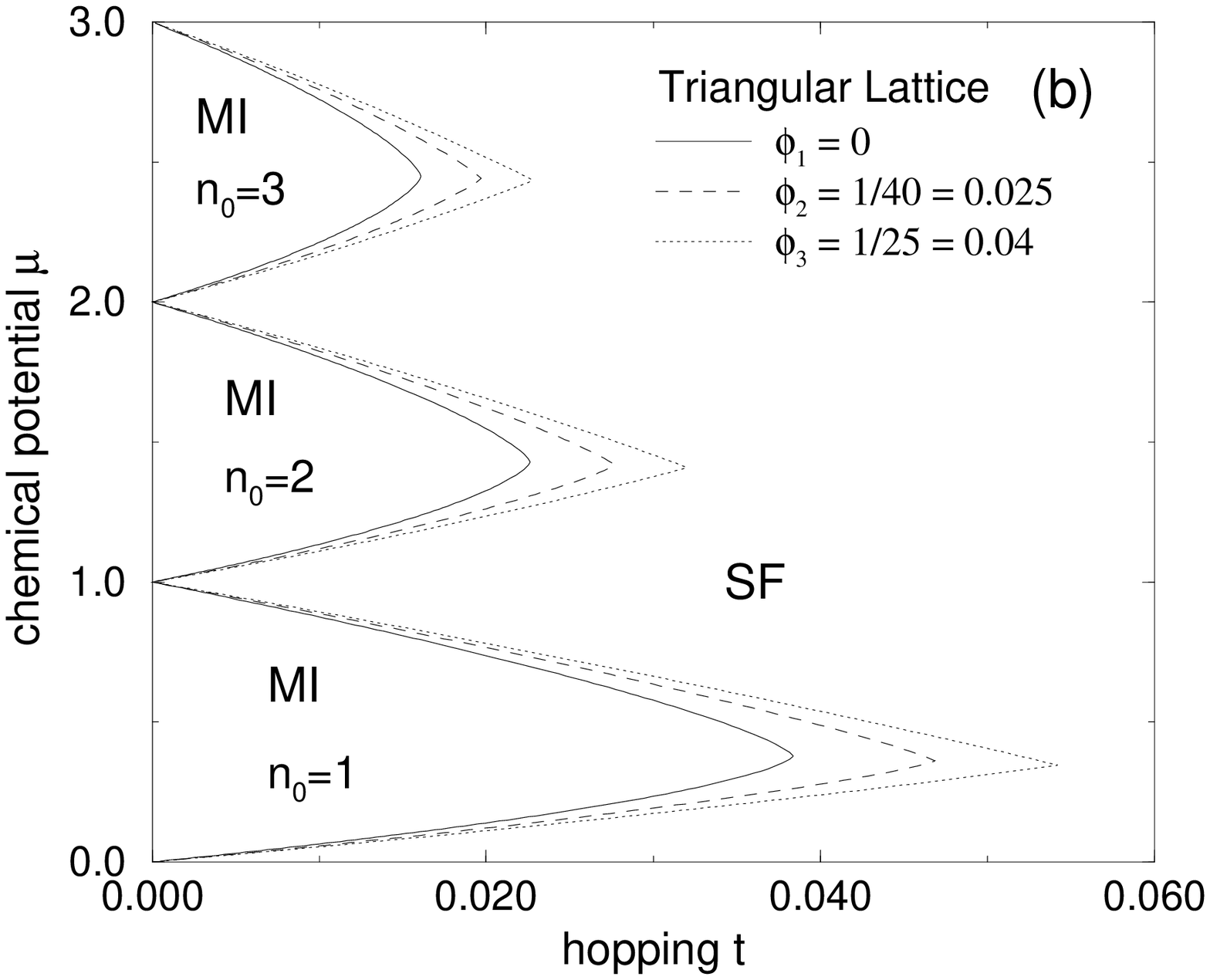,
                  height=\hoehe,width=\breite}}
        }
\end{picture}
\caption[]
        {Phase diagram for the square (a) and the triangular (b) lattice in
          relatively small magnetic fields using the Pade analysis.}
\label{phasesqtri}
\end{figure}

Fig.~\ref{phasesqtri} presents the phase diagram of the first three lobes
for the square lattice \ref{phasesqtri}(a) and the triangular lattice
\ref{phasesqtri}(b) in the low magnetic-field region.  The
incompressible, Mott-insulating phase grows in size when the magnetic
field increases from zero.  This shows, as expected, the localizing
effect of the magnetic field on the itinerant bosons.  It appears
that the shape of the lobes changes from a simple
power-law dependence at the tip to something else.  There are three
possibilities for the shape of the phase lobe at the tip: (i) it remains
power-law-like with an exponent that may depend on the magnetic field;
(ii) it has a more exotic ``cusp-like'' dependence as seen in the
Kosterlitz-Thouless transition; or (iii) it has a discontinuous change
in slope at the tip, and thereby has a ``first-order-like'' shape
corresponding to the crossing of the two curves representing the upper
and lower phase lobes, with no ``critical'' behavior at the tip (this
last result corresponds to $z\nu=1$).  There is theoretical evidence
in favor of this last conclusion, which is a similar behavior to what
happens to the MI phase in the presence of disorder, but here the
explanation is different\cite{ma-private}.  If $\phi=p/q$ is expressed
in lowest terms, then the order parameter requires $q$ components to
describe it.  As $q$ increases, it is more likely that the transition
is first-order rather than second-order, and hence one would expect
that the tip has a slope discontinuity immediately upon the
introduction of the magnetic field, since small fields correspond to
large values of $q$.  The Pade analysis given above assumes that (i)
holds, but as $z\nu$ approaches 1 for larger fields, the system
crosses over to the behavior of (iii) unless something more exotic
like (ii) intervenes.  We cannot distinguish between scenarios (i) or
(iii) with the low-order expansions presented here, but our results
suggest scenario (iii) is correct, because the exponent $z\nu$ rapidly
approaches 1 as the field is increased.  Clearly further
theoretical analysis is needed to decide this issue.

For larger magnetic fields, we find that the two boundaries of the
Mott phase lobe no
longer cross at a critical value of $t$, but rather they ``repel''
each other.  This indicates that the tip of the MI phase lobe has
moved out to such a large value of $t$ that the perturbation theory
needs to be carried out to higher order to achieve proper convergence.
Such an exercise has been already carried out through a linked-cluster
expansion for the fully frustrated case ($\phi=1/2$) on a triangular
lattice\cite{monien98}.  The calculation from 3rd to 11th order
converges toward a ``first-order'' transition $(z\nu =1)$ at
$t_{crit}\approx 0.06$ which yields a much larger insulating phase
regime than with no magnetic field where $t_{crit}\approx 0.038$.  A
comparable experimental result is found\cite{mooij} where the insulator to
superfluid transitions on Josephson-junction arrays were studied by
changing the ratio of the Josephson coupling energy $E_J$ and the
charging energy $E_C$. Since $E_J$ is the energy scale for the
superconducting coupling between the islands and $E_C$ is the scale
for the interaction between the charge carriers, they can be related to the
hopping energy $t$ and the on-site repulsion $U$, respectively.  The
results for small temperatures can be extrapolated to $T=0$ and show a
transition at larger $E_{J}/E_{C}$ for the $\phi=1/2$ case than for
the $\phi=0$ case, which qualitatively matches with our calculations (since
$t_{crit}/U$ increases with magnetic field).

The experimental data can be analyzed in one of two dual pictures with the Bose
Hubbard model.  In the first view, the bosons are the Cooper pairs on
each island, and the main effect of the magnetic field is to modify the 
hopping matrix.  Vortices appear in this case as supercurrent loops in the 
system.  While the density of vortices increases as the magnetic field is 
modified, the density of Cooper pairs remains essentially constant over the 
low field ranges (on the order of Gauss)  explored in the experiments.  
The second picture employs the duality transformation that views the vortices 
themselves as the bosonic  particles\cite{duality}.  The mobility of the 
vortices is determined by $E_C$ and their interaction by $E_J$, so the roles of 
those parameters are reversed in this case.  The magnetic field then takes on 
the role of the chemical potential.  We will not employ this second picture 
here, which has been used to evaluate the Mott insulating phase of the vortices
in the quasi-one-dimensional Josephson junction arrays\cite{mooij_1d}.

In the following we compare small-field results. As shown in
Fig.~\ref{phasesqtri} the MI phase area and the location of the tip of
the lobe increase with increasing magnetic field up to about
$\phi=0.1$ (for the square lattice). The measurements of \cite{mooij} show
$(E_J/E_C)_{crit}\approx 0.59$ and $0.83$
for $\phi=0$ and $0.1$ respectively, which is similar
to our results with $t_{crit}$ increasing by about 45\% from 0.057 to
0.083 as $\phi$ increases from 0 to 0.1.
In addition, the tip location for larger
field strengths saturates in the experiments,
which is a result that we cannot confirm, because our
analysis through third-order in $t$ fails when $t_{crit}$ becomes too large.

\begin{figure}[t]
\setlength{\unitlength}{1mm}
\def\hoehe{64mm}
\def\breite{64mm}
\vspace{1.0cm}
\begin{picture}(60,50)(0,0)
\put(21,0){
        \makebox{ \epsfig{file=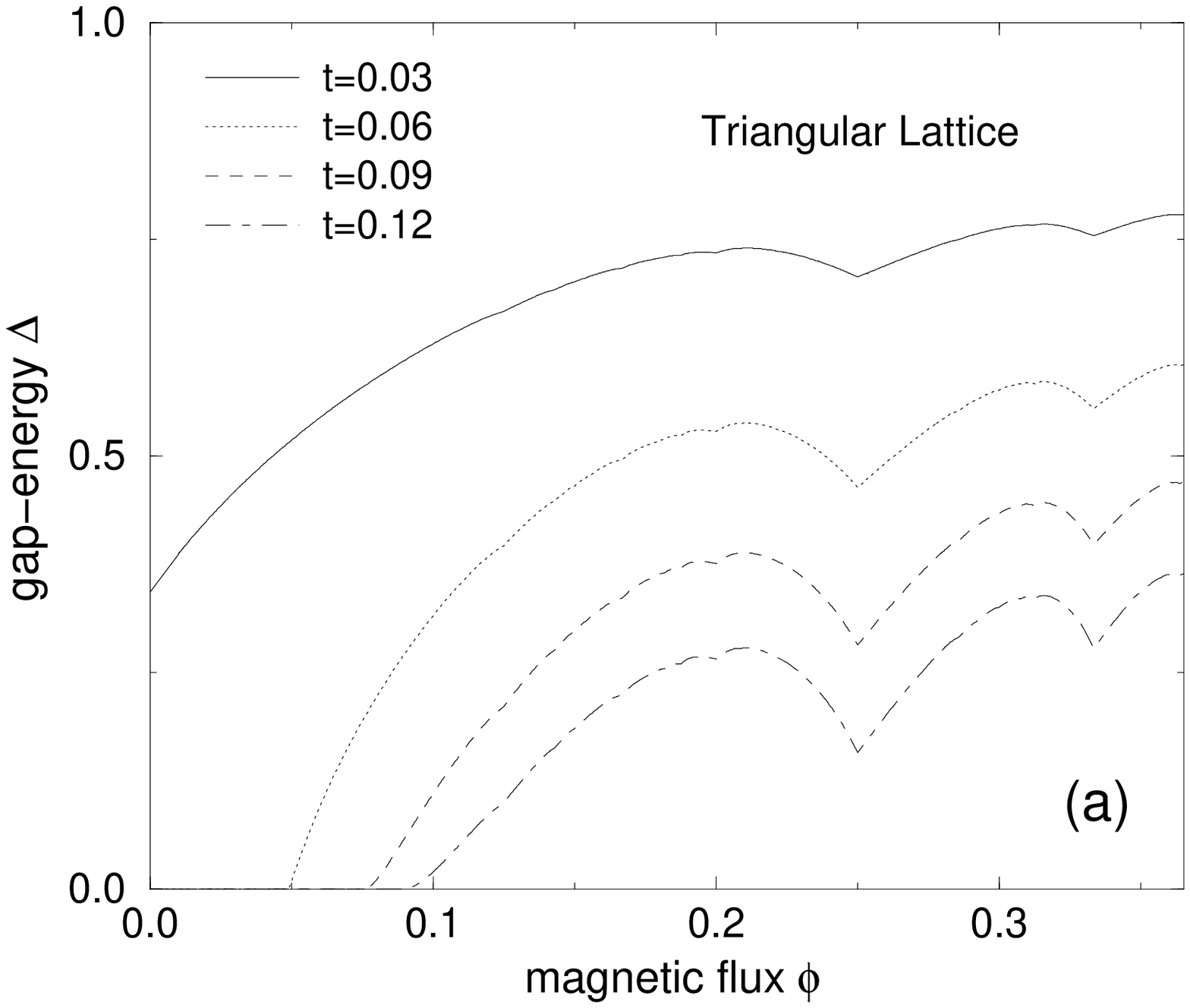,
                  height=\hoehe,width=\breite}}
        }
\put(86,0){
        \makebox{ \epsfig{file=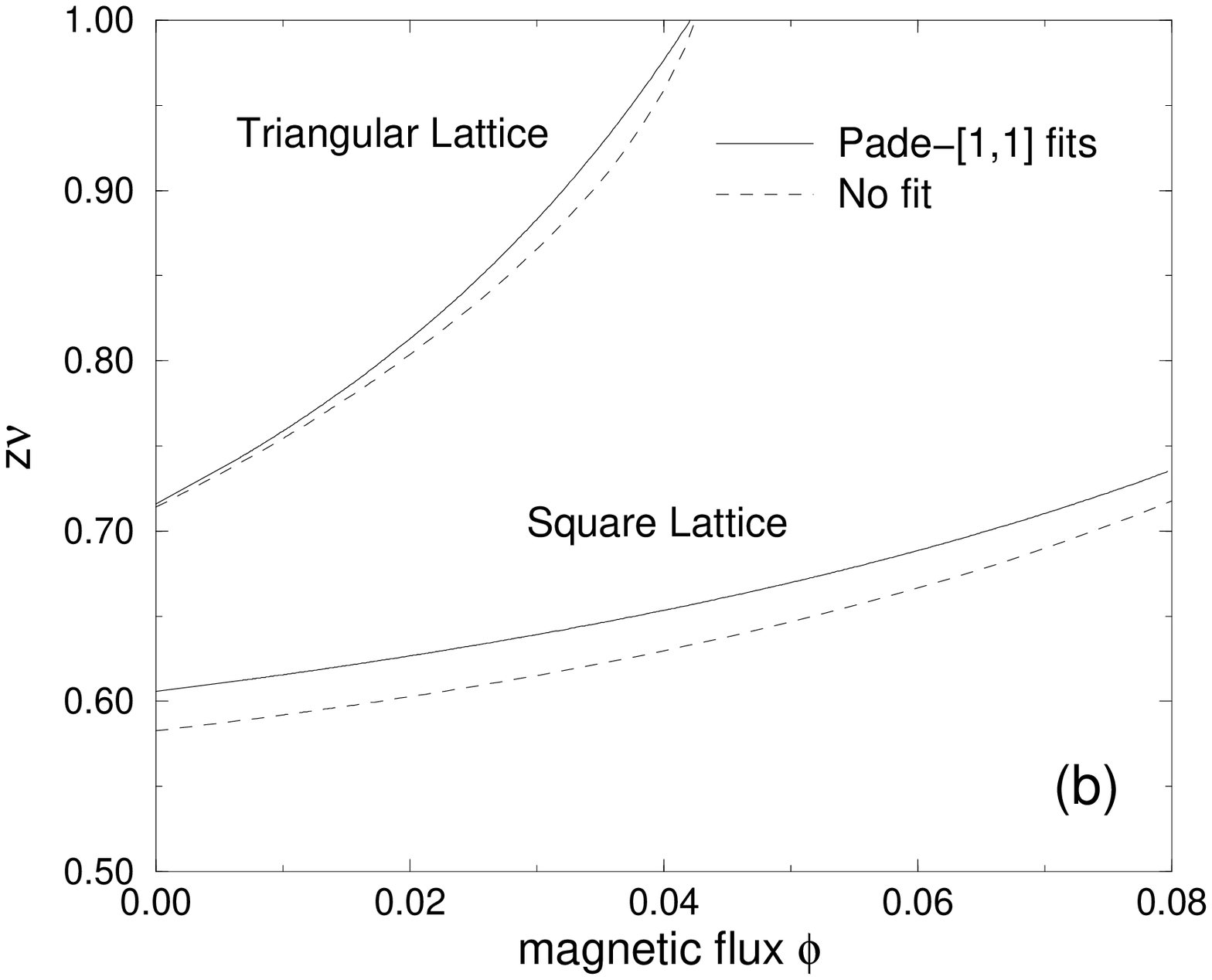,
                  height=\hoehe,width=\breite}}
        }
\end{picture}
\caption[]
{ (a) Dependence of the gap energy (for fixed hopping) on the magnetic
  field. Notice the kinks that develop at commensurate magnetic fields
  ($\phi=1/3$ or 1/4). The curves are cut off at $\phi\approx 0.35$ 
  because the perturbation theory fails for larger values of $\phi$.
  (b) Dependence of the dynamical critcal
  exponent $z\nu$ on magnetic field when the truncated expansions, or
  a Pade approximant, are fit to the power-law behavior at the tip of
  the lobe.  Notice that $z\nu$ appears to be field-dependent and
  rapidly approaches 1.}
\label{exponent}
\label{deltamutri}
\end{figure}

In Fig.~\ref{deltamutri}(a) we show the evolution of the
excitation-gap energies $\Delta(t)$ for fixed hopping $t$ as a function of
magnetic field. Initially the gap energy increases which indicates the
increasing stability of the MI phase in a weak magnetic field. For
larger magnetic fields, we see dips in the gap energy around rational
magnetic fluxes such as $\phi=1/3$ or $1/4$ (which can be reliably
calculated for small values of $t$).  A similar commensurable
structure is found in experiments that measure the zero-bias
resistance $(R_0)$ of a Josephson-junction array
in a magnetic field\cite{mooij}.  A small
$R_0$ indicates proximity to the SF phase.  At rational $\phi$ the
vortices form a lattice which is commensurable with the Josephson-junction
array and favors a pinning of the vortices, leading to a decrease of the
zero-bias resistance. Qualitatively, we see the same behavior in the excitation
gap energies and the zero-bias resistance.
Notice how the dips track closely with the dips seen in the minimal eigenvalue 
of the hopping matrix as shown in Fig.~\ref{butterfly}.  This behavior explains
the dips seen in the experimental data on the triangular lattice
that are most prevalent at 1/2, 
followed by 1/4, 1/3, 3/8, and so on.  The experimental data\cite{mooij} also 
shows that the system has four different regions of superconducting stability 
for one value of $t$, centered at 0, 1/2, 1/4 and 1/3.  While our results in 
Fig.~\ref{deltamutri}(a) only show superconductivity around $\phi=0$
[where $\Delta(t)=0$], we can 
see that if $t$ was increased, and the calculations carried out to higher order,
it is likely that we would see additional superconducting regions appearing 
(we believe first around 1/2, followed by 1/4 and 1/3).  

Finally we study the magnetic-field dependence of the dynamical
critical exponent $z\nu$ (Eq.~\ref{znuformel}) for small fields in
Fig.~\ref{exponent}(b).  The BHM in zero magnetic field can be mapped
onto a three-dimensional XY model which has $z\nu=0.67$ 
independent of the lattice structure\cite{guillou}.  
By using the Pade analysis on our third-order
expansion, we obtain $z\nu=0.61$ for the square lattice and
$z\nu=0.72$ for the triangular lattice at $\phi=0$.  A Pade analysis
of a tenth-order expansion\cite{monien98} yields $z\nu=0.69$ for both
lattice types, which shows the convergence of higher-order
calculations in our method.  As seen in Fig.~\ref{deltamutri}, the
dynamical exponent appears to increase as the magnetic field
increases.  It is, however, difficult to conclude whether $z\nu$
remains equal to it's zero-field value, increases with magnetic field,
or immediately jumps to one upon the onset of a magnetic field, solely
on the basis of this third-order analysis.
But, the fact that the most likely point to have $z\nu<1$ is $\phi=1/2$
(because it requires only two components for the order parameter), and a 
higher-order expansion predicts $z\nu=1$ there, 
leads us to conjecture that $z\nu=1$ for all nonzero magnetic fields.  This 
latter result is supported by Monte Carlo data on the antiferromagnetic
XY model for stacked 
triangular planes, which have a weakly first-order transition rather than a 
critical point\cite{stackedxy}.

\section{Conclusions}

In conclusion, we applied a strong-coupling $t/U$-expansion (up to
third order) to study the insulator-superfluid phase transitions of
the two-dimensional Bose-Hubbard model under the influence of a
magnetic field.  
Although our analysis is limited, we are able to produce reasonable
results that both agree with physical intuition and with experiments.
We found that the Mott insulating phase enlarges with an increasing
magnetic field. This is explained by the localizing effect of the
magnetic field on the moving bosons. Qualitative agreement in the
increase of the critical hopping energy $t_{crit}$ was found with
experimental results on Josephson junction arrays\cite{mooij}.  For
small magnetic fields, we find a power-law behavior of the gap energy
close to the critical point ($t_{crit},\mu_{crit}$) with a dynamical
critical exponent that increases with $\phi$.
For larger magnetic fields, we find a repulsion of the two phase
boundaries, which indicates a change from the power-law like
behavior to either Kosterlitz-Thouless behavior or to a ``first-order''
transition.  Our results are also consistent with the ``critical point''
immediately changing to a ``first-order'' discontinuous change in the
slope of the phase boundaries as the magnetic field is turned on.
We found that the gap energies for small fixed hopping
and variable magnetic field illustrate commensurability effects for
rational fluxes which is also seen in Josephson-junction array
experiments.  More work is
needed to understand the change in character of the insulator to
superfluid phase transition as a magnetic field is introduced: does
the system have a field-dependent power-law dependence which crosses
over to a ``first-order'' (or more exotic cusp-like shape) as the
field increases, or does it immediately become ``first-order'' in a
field.  Higher-order calculations are needed to decide this issue. It would also
be interesting to extend the scaling analysis of the BHM to include its
behavior in an external magnetic field.

\acknowledgments 

We would like to thank M. Ma for useful and interesting discussions.
J.K.F. acknowledges support from an ONR YIP grant N000149610828 and
from the Petroleum Research Fund administered by the American Chemical
Society (ACS-PRF 29623-GB6).

\end{document}